\documentclass[%
 reprint,
 amsmath,amssymb,
 aps,
pra]{revtex4-1}

\usepackage{graphicx}
\usepackage{hyperref}

\begin{document}         



\title{X-ray optics and beam characterisation using random modulation: Experiments}


\author{Sebastien Berujon}
\email{xebastien@berujon.org}
\affiliation{European Synchrotron Radiation Facility, CS 40220, F-38043 Grenoble cedex 9, France}
\author{Ruxandra Cojocaru}
\affiliation{European Synchrotron Radiation Facility, CS 40220, F-38043 Grenoble cedex 9, France}
\author{Pierre Piault}
\affiliation{European Synchrotron Radiation Facility, CS 40220, F-38043 Grenoble cedex 9, France}
\author{Rafael~Celestre}
\affiliation{European Synchrotron Radiation Facility, CS 40220, F-38043 Grenoble cedex 9, France}
\author{Thomas Roth}
\affiliation{European Synchrotron Radiation Facility, CS 40220, F-38043 Grenoble cedex 9, France}
\author{Raymond Barrett}
\affiliation{European Synchrotron Radiation Facility, CS 40220, F-38043 Grenoble cedex 9, France}
\author{Eric Ziegler}
\affiliation{European Synchrotron Radiation Facility, CS 40220, F-38043 Grenoble cedex 9, France}




\date{\today}

\begin{abstract}
In a previous paper \cite{berujon2019a}, we reviewed theoretically some of the available processing schemes for X-ray wavefront sensing based on random modulation. We here show experimental applications of the technique for characterising both refractive and reflective optical components. These fast and accurate X-ray at-wavelength metrology methods can assist the manufacture of X-ray optics that transport X-ray beams with minimum amount of wavefront distortion. We also recall how such methods can facilitate online optimization of active optics.
\end{abstract}

\maketitle

\section{Introduction}

In our theoretical paper \cite{berujon2019a}, we recalled the available X-ray near-field speckle-based methods for at-wavelength metrology and their associated processing schemes. These speckle-based methods make use of high frequency wavefront intensity random modulations whose small features act as markers permitting to infer the rays' trajectory. Since within the X-ray regime most of the available beams are only partially coherent at best, the observed X-ray speckle patterns do not fully develop as a result of an interference process but rather as the product of a mix of phase and amplitude modulation. The speckle generator needed to infer the phase distortion of the beam can be regarded as a random mask that has the possibility to move perpendicular to the direction of the beam in order to provide additional or redundant information.

The speckle method, although numerically more sophisticated than other methods, can be easily implemented on a beamline. As a matter of fact, the method solely requires a speckle generator that is often very easy to source (e.g. a piece of sandpaper) and the distances involved are not very strict, permitting to quickly and easily fit into the instrumentation on many beamlines. However, to achieve an accuracy on the wavefront gradient in the order of one nanoradian, a (piezoelectric) nano-positioner is required.

The speckle-based methods are namely: X-ray speckle tracking (XST) \cite{berujon2012prl}, X-ray speckle vector tracking (XSVT) \cite{berujon2015pra}, X-ray speckle scanning (XSS) \cite{berujon2012pra} and a hybrid processing scheme, less simple to implement, but also advantageous in several aspects. In the XST approach the beam phase is inferred using a dual-image correlation-based technique that enables single-pulse wavefront metrology \cite{berujon2015josr}. The other methods require micrometer-accurate scanning of a speckle generator object, eventually permitting to achieve nanoradian sensitivity and pixel-size resolutions \cite{berujon2015pra}.

In this paper we demonstrate the performances of the different speckle techniques by providing their corresponding implementations at a synchrotron beamline. We illustrate the potential of each method in a variety of at-wavelength metrology situations, using both refractive and reflective optics. Refer to the first article of this dyad for more details regarding the numerical processing algorithms used herein \cite{berujon2019a}.

At the ESRF, and more specifically at its Instrumentation Facility, beamline BM05 \cite{ziegler2004}, speckle-tracking methods are regularly applied for the at-wavelength characterisation of new optical components and for development purposes. In this article, we present implementations of the previously presented near-field speckle schemes with experimental applications for beam, lens and mirror characterisation.

Although the main focus of this work is on wavefront sensing and metrology for the characterisation of X-ray optical components, one can envisage the applicability of the proposed methods to visible light optics as suggested by recent works \cite{berto2017}.

\section{Experimental aspects}

Although the requirements placed on the X-ray beam for the implementation of speckle-based techniques are rather minor, a couple of factors can induce systematic errors. One should try to independently minimize the errors induced by the beam, the detector and the speckle generator. In the following, we provide several recommendations to help the reader in achieving his own implementation of the techniques.

\subsection{Data processing}

The data presented here were processed using home-made routines coded in MATLAB. A non-supported version of some of these codes may be obtained from the authors upon request, whilst another more robust Python implementation is readily available at \texttt{https://gitlab.esrf.fr/cojocaru/swarp}, providing routines for the calculation of detector distortion and for wavefront analysis of the XST implementation schemes.

Beyond the code necessary for recovering the beam phase, mentioned above, additional processing routines must be integrated in order to extract additional information such as phase error, the thickness of a transmission optics or a wavefront modal decomposition. As an example, see the processing steps for an X-ray lens at \texttt{https://gitlab.esrf.fr/cojocaru}.

\subsection{Detector distortion}

The detector can introduce systematic errors in the measurements. The main source of such errors is the distortion induced by the optical system used for recording the high-resolution images. For hard X-rays, high-resolution imaging detectors use a scintillator to convert X-rays to visible light. The visible light image is then recorded by a digital camera through a magnifying optical system. The presence of lenses and of a microscope objective are the main sources of image distortion. These systematic errors must be corrected for, as it is also the case for grating-based methods~\cite{inoue2018}.

\begin{figure}
\includegraphics{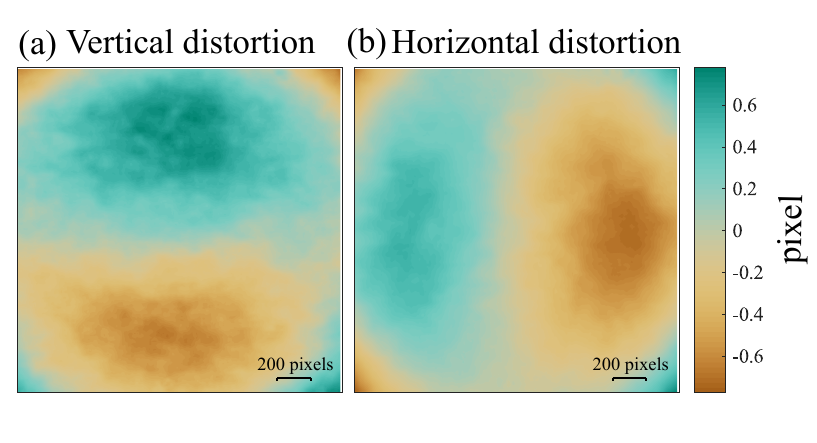}
\caption{(a) Vertical and (b) horizontal detector distortions. The large stripe on the left corresponds to a mark on the Kapton window protecting the scintillator and is responsible for a local subpixel amplitude distortion. \label{fig:detdisto}}
\end{figure}

We use the technique described in~\citet{berujon2015josr} to characterise the optical distortion of the imaging system at the beginning of each of our experiments using a 2D mesh scan of the detector positions perpendicular to the X-ray beam, consisting of 25 points placed on a square grid and located 20~$\mu$m apart. Figure~\ref{fig:detdisto} displays the typical distortion of one of our X-ray imaging systems consisting of a CMOS PCO Edge camera, a thin LSO:Tb scintillator with 10~$\mu$m thickness and a magnifying optics that eventually leads to an effective pixel size of $s_{pix} = 0.615~\mu$m. In this case the peak-to-valley amplitude of the distortion is on the order of two pixels but can be much larger when employing detectors with larger chips and/or a microscope magnifying objective of lesser quality.


The calculated distortion is not used to correct all the recorded raw images, but only the calculated gradient maps, using a two-dimensional bicubic interpolation. In the case of XSS, as the processing scheme operates pixel by pixel, no systematic errors from the detector are introduced in the angular measurement. In the other cases where neighboring pixels are included in the processing algorithm via a multi-pixel window function, the local detector distortion from one pixel to the neighboring ones is so small ($ < 10^{-3}$ pixel per pixel) that its effect on the displacement vector is negligible. The only case where the raw images must be corrected is for the XST and XSVT methods used in absolute mode, since the same part of the beam will hit different areas of the detector for the two images. In such a case, the distortion between these two detector areas becomes a source of error for the calculation of the angular deflection generated by the phase gradient. 

Further processing like the fitting of polynomials to the calculated wavefront error does require the distortion correction, because global (full detector frame) maps are considered in such an analysis.

\subsection{Choice of a speckle generator material}

The choice of the speckle generator material is not very restrictive and only a few practical guidelines can be given. One may want to empirically test a few scattering objects to find the one best suited for a certain experiment.

The criteria defining a good speckle generator object are the visibility of the speckle grains in the images and their size. In practice the dimension of the grains should be selected so as to cover a few pixels of the image ($<$10 pixels) and with a minimum contrast of 0.1, where contrast can be defined as the standard deviation over the average intensity. Depending on the photon energy and on the detector used, various choices can prove to be suitable, such as granular materials, sandpapers, filters (e.g. cellulose with micrometer pore sizes), used individually or in stacks. While the transverse coherence of the beam will help increase the speckle contrast, the absorption of the material can also be used to generate speckle grains. In that case though, care should be taken to select a speckle membrane generator with an acceptable total absorption, that is stable in time, can handle the photon flux and does not diffract too much at larger scattering angles.

\section{Metrology applications}

The following metrology examples aim to demonstrate the validity of the speckle-based phase-sensing methods in various scenarios. As mentioned before, beyond the code necessary for phase retrieval from the speckle images, one should also consider subsequent analysis needs, usually involving the use of dedicated routines, specifically written for optics metrology using an X-ray beam. For instance, the recovery of the shape error of a mirror requires the removal of the best surface, which depends on the design parameters of the reflective optics. In contrast, lens metrology will often call for polynomial wavefront decomposition in order to analyse the type of aberrations produced in the system. Such processing can represent a large part of the overall numerical treatment and require user input, for example when defining the mask to be used when analysing a lens.

The metrology experiments presented here were conducted at the beamline BM05 of the ESRF~\cite{ziegler2004}. This beamline is a multipurpose facility equipped with a characterisation platform devoted to the measurements of various optical components. The photons are produced by a 0.85 T dipole with electrons circulating in the ESRF storage ring with a nominal energy of E = 6.03 GeV. The X-ray spectrum accessible at the beamline is continuous and determined by the accelerator parameters. The critical photon energy of the bending magnet radiation is 19.2~keV.

\subsection{Beam diagnostics}

\subsubsection{Absolute beam state}

This first example aims to demonstrate the use of the XSVT method for the measurement of the absolute beam wavefront at a position of interest along the beamline.

\begin{figure}
\includegraphics[width=8.5cm]{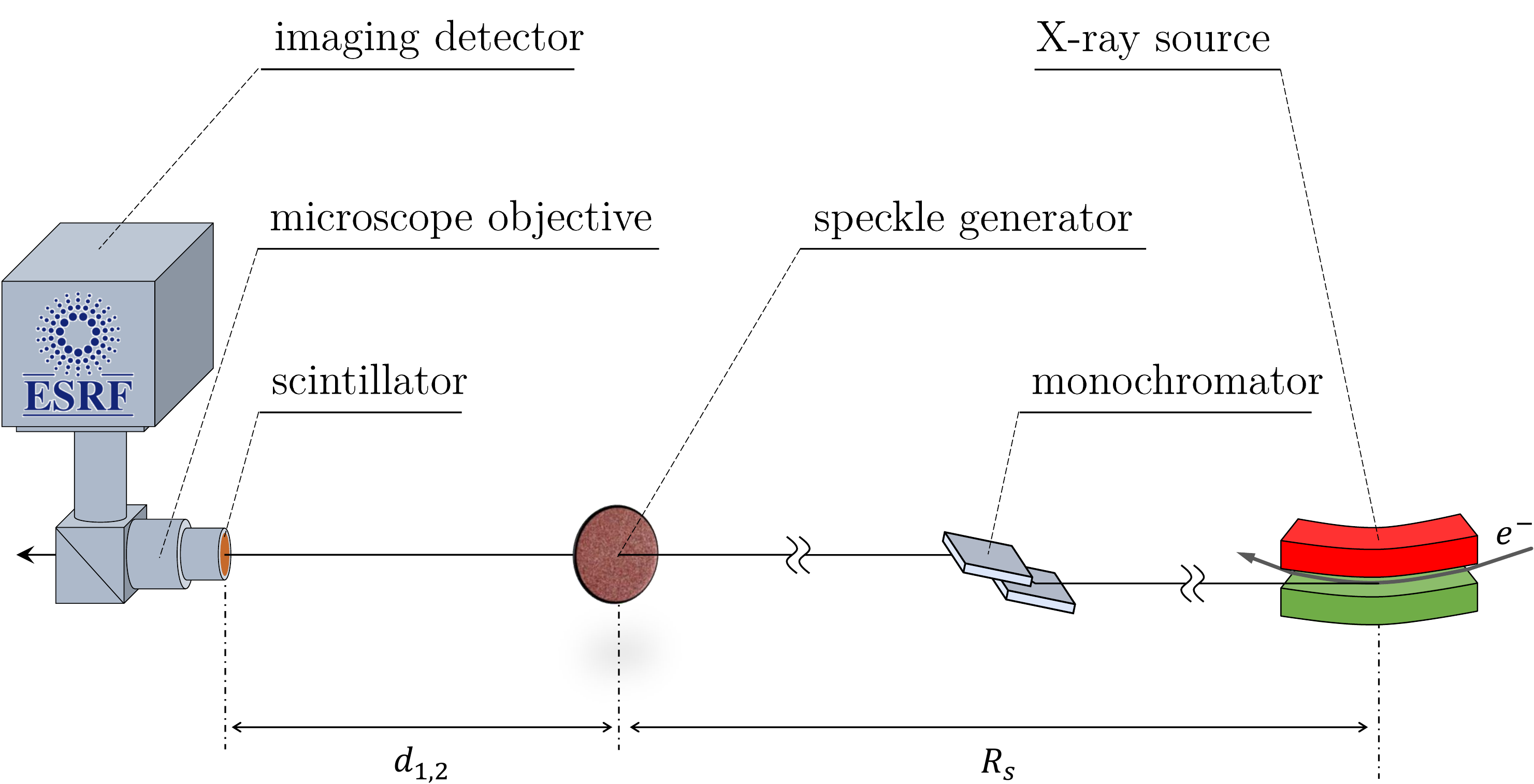}
\caption{Sketch of the experimental configuration used to measure the X-ray wavefront state and stability of the ESRF BM05 beam after reflection on a double crystal or multilayer monochromator. \label{fig:sketchnosample}}
\end{figure}

The beam measured is that of BM05, ESRF, when using a vertically deflecting multilayer monochromator with the energy set to $E\!=\!17$~keV. The multilayer monochromator was designed with flat mirrors preserving the natural divergence of the source upon reflection of the X-rays. The experimental configuration is sketched in Fig. \ref{fig:sketchnosample} where the detector was moved along the beam axis between different data acquisitions. Two sets of 100 images at 100 well-defined (but not equally spaced) transverse positions of the speckle membrane were collected sequentially in two propagation planes separated by $\Delta d=d_2 - d_1=200$~mm and located at approximately $R_S=40.5$~m from the source. The detector was a FReLoN camera coupled to an magnifying optics imaging a 10~$\mu$m thick LSO:Tb scintillator, rendering an effective pixel size of $p_{ix}=0.78~\mu$m. Since XSVT has a sensitivity proportional to the detector pixel size, a high resolution imaging detector is strongly recommended when implementing this technique in order to maximize its performance.

The XSVT approach was preferred here to the other methods, such as XSS in the self-correlation mode which could have also been an option. The XSS scheme presents the alternative advantage of not requiring the translation of the detector during the data collection, just that of the speckle generator. This can prove essential when space available for beam diagnostics is limited. However, XSS also has some disadvantages that must be taken into account, like for instance its sensitivity to the local curvature of the beam. The use of XSS requires three separate steps of 2D numerical integration to reconstruct the beam's wavefront versus the single integration step required by XSVT.
In our present case, the use of a high resolution detector with XSVT helps to gain angular resolution since it is proportional to the detector's pixel size.

\begin{figure}
\includegraphics[width=8.5cm]{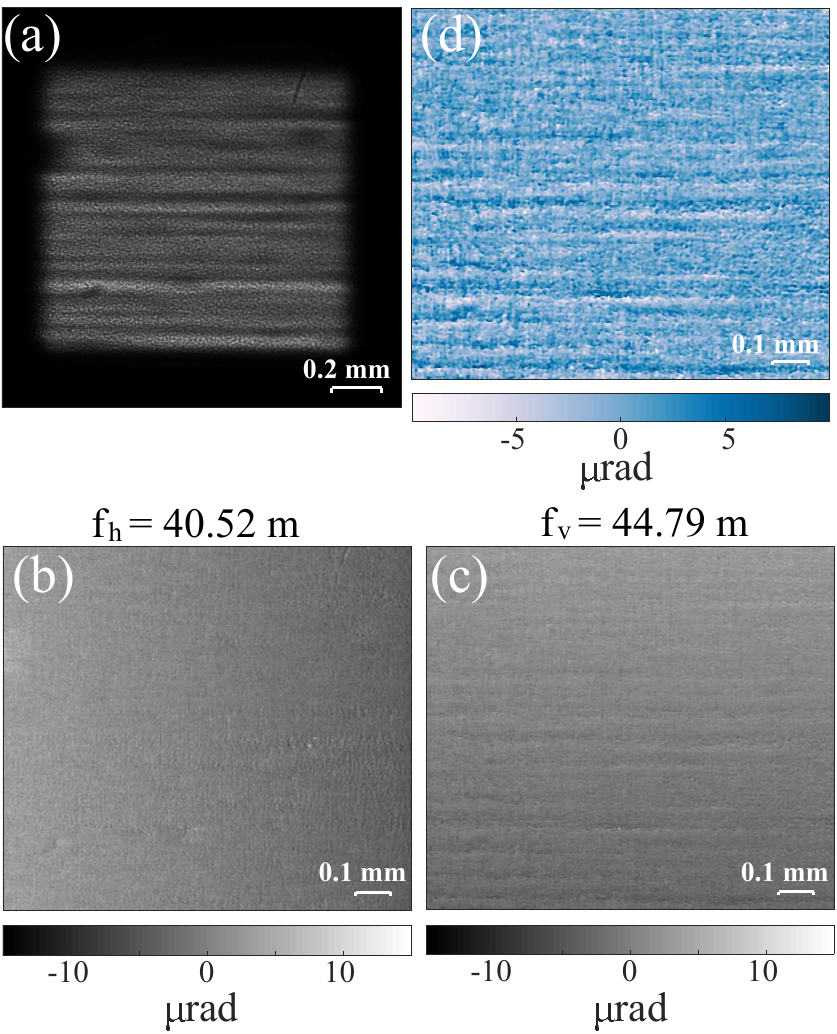}
\caption{Absolute wavefront as measured with speckle vector tracking. (a) First raw image. (b-c) Horizontal and vertical wavefront gradients. These gradients are compatible with a curved wavefront generated $f_h$=40.5\,m or $f_v$=44.8\,m away. (d) Vertical wavefront gradient error. \label{fig:absvect}}
\end{figure}

Figure~\ref{fig:absvect} shows in (a) the speckles generated by the speckle membrane and the intensity fringes generated by the multilayer monochromator upon beam reflection. These are noticeable in the vertical wavefront gradient, shown in (c), and correspond to angular aberrations of the order of a few tens to hundreds of nanoradians as we shall see further on. An important parameter that can be extracted from this metrology is the slight astigmatism of the beam. While in the horizontal direction (transverse to the monochromator double-bounce reflection in the vertical direction) the extracted source distance perfectly matches the physical distance from the source to the detector, a discrepancy is observed in the vertical direction. The virtual vertical source distance calculated is $f_v = 44.8~$m compared to an expected value of $f=40$~m. This aberration is introduced by the monochromator, perhaps due to surface errors of the mirrors or possibly due to a thermal bump generated on the first monochromator mirror.

\subsubsection{Beam stability measurements}

The beam stability of a beamline can be monitored using XST, using sequences of images recorded with a fast time sampling resolution. XST offers a good alternative for stability measurements since only a single image is needed to compare with a reference one, the method hence allowing to track dynamic processes.

The example presented is again for the ESRF beamline BM05 after monochromatization at $E\!=\!17$~keV using the beamline's double-crystal Si(111) monochromator, with the setup presented in Fig. \ref{fig:sketchnosample}.

In this experiment, the achievable time resolution limit is fixed by the X-ray flux density and by the minimum number of counts necessary on the detector for reliable tracking. The setup consisted solely of the static speckle generator and a PCO Edge 4.2 camera continuously acquiring images of the beam with an exposure time of 0.0125~second per frame (80 Hz). The indirect magnifying optics (the X-ray image being converted to visible light through a LSO:Tb scintillator) led to an effective pixel size of $p_{ix}=1.6~\mu$m. The distance from the speckle generator to the detector was $d = 840$~mm, and $R_S=40.5$~m remained unchanged.
\begin{figure}
\includegraphics[width=8.5cm]{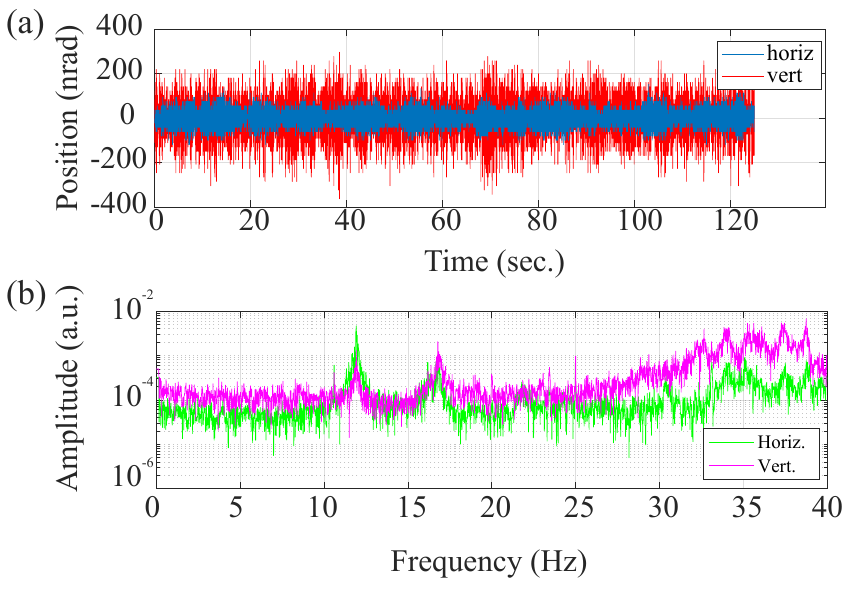}
\caption{(a) Angular position of the beam tracked from image to image using the XST method. (b) Frequency decomposition showing the vibration modes. \label{fig:beamstability}}
\end{figure}

Figure~\ref{fig:beamstability}~(a) shows the measured angular positions of the beam as obtained by tracking the same large subset of pixels - and thus the speckles contained in it - across all the images collected over time. The angular positions are centered around the average position of the beam. Figure~\ref{fig:beamstability}~(b) presents the time spectral decomposition of the angular beam positions. From these graphs, one can identify vibration modes, already known at the ESRF, around 35~Hz but also peaks near 12~Hz and 18~Hz originating from flow induced vibrations coming from the monochromator cooling system. This application illustrates the potential of the method for beamline optimization and design.

The use of the XST based method for monitoring the beam stability presents the advantage of being much more photon efficient with respect to alternative approaches that use a pinhole or focusing optics. Indeed, there is no or very little absorption in the monitoring device and many pixels are used to track the angular direction of the beam whilst only a few pixels are used with the previously mentioned competing methods. The improved photon statistics offers a better performance in terms of noise and sensitivity. Moreover, the XST method can be easily applied at higher energies where the pinhole and optics based methods become challenging to implement.

\subsection{Reflective optics}

\subsubsection{2D mirror surface characterisation\label{sec:2dmirror}}

Online mirror metrology using the XSS technique was demonstrated in \citet{berujon2014oe}. A simplified and truncated two-dimensional approach of the technique was later published in \citet{wang2015josr} which can be effective for flat mirrors. In the following we demonstrate a more exact way of recovering the mirror shape for a strongly focusing mirror from 2D mesh scans of the speckle generator using the 2D absolute XSS method with self-correlation (see Sec. 2.2.5 of the theoretical paper \cite{berujon2019a}).

\begin{figure}
\includegraphics[width=8.5cm]{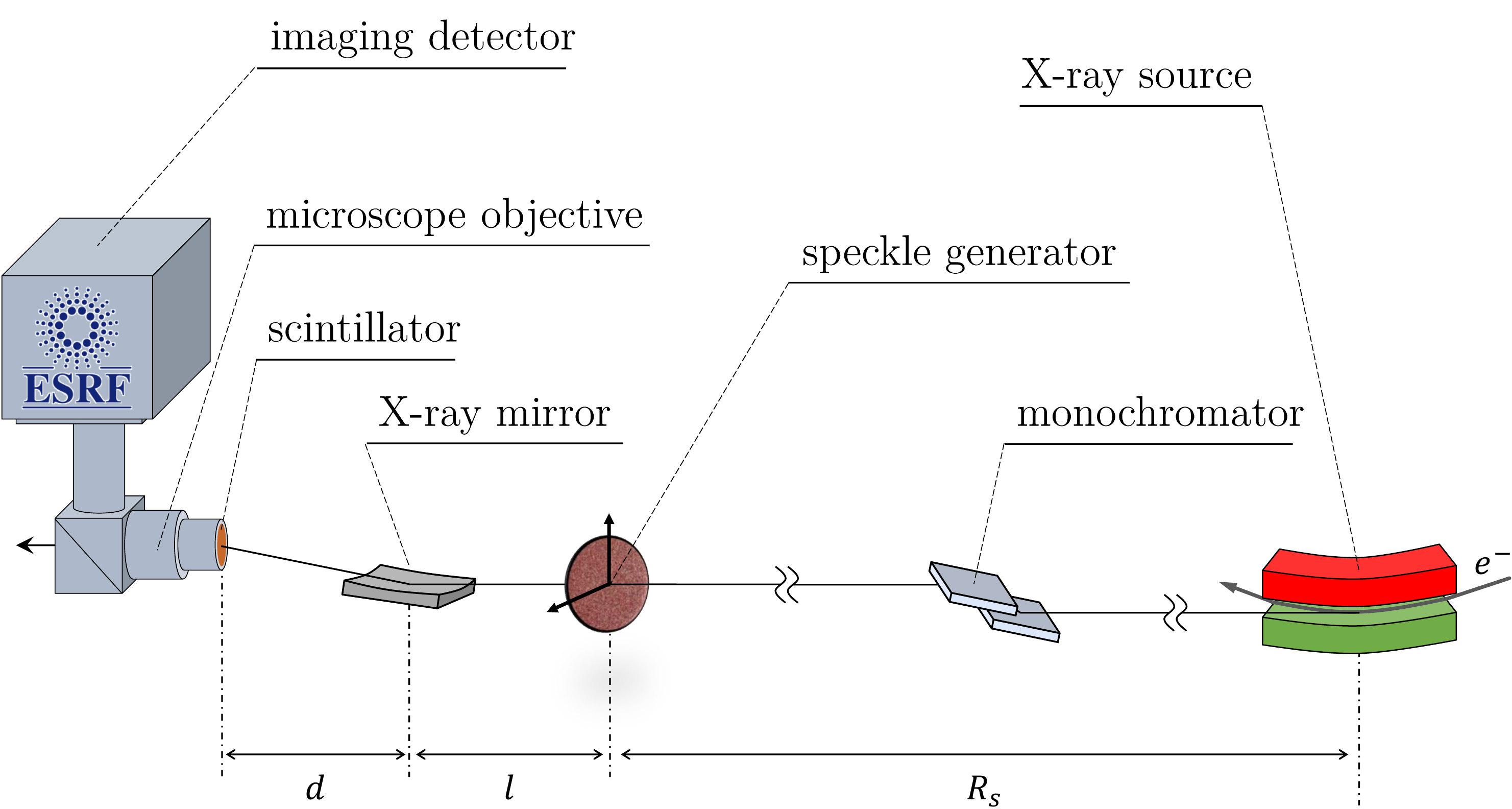}
\caption{Setup used for the reflective optics characterisations. \label{fig:setupreflect}}
\end{figure}

The XSS approach revealed very early its efficiency for the characterisation of reflective optics \cite{berujonphd}. Indeed, when measuring a strongly focusing X-ray optics, the magnification generated by the investigated mirror (or alternatively by a lens, as seen in Sec. \ref{sec:strefop}) permits to use smaller speckle generator steps since they get enlarged by propagation and become detectable on the detector. That leads to the detection of smaller displacement vectors, thus achieving a higher accuracy in the method (see Table of \cite{berujon2019a}). Combined with a propagation distance $d$ of several meters, the ultimate angular resolution can reach the order of a single nanoradian.

\begin{figure}
\includegraphics[width=8.5cm]{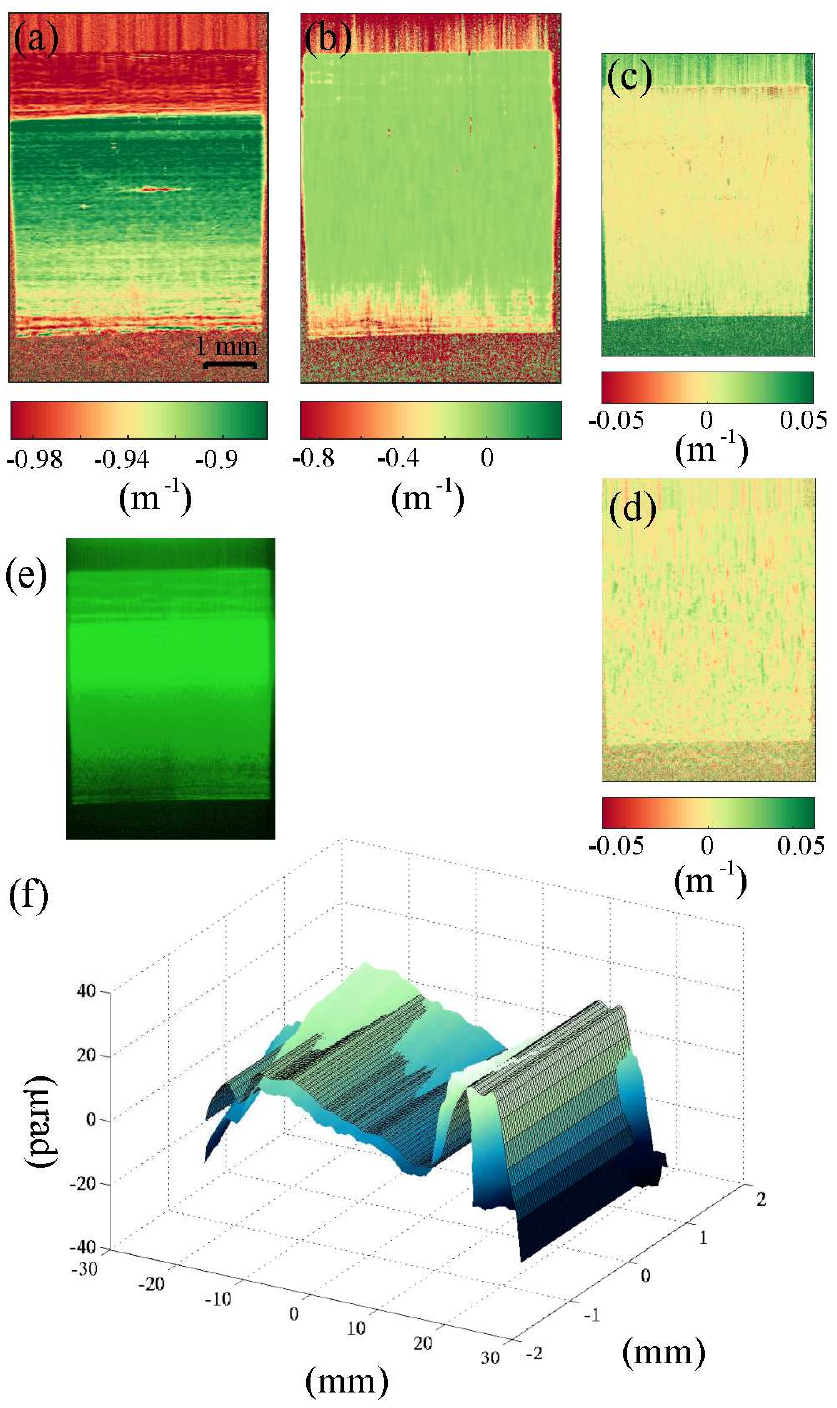}
\caption{Curvature field (a) $\partial^2 W/\partial x^2$ (b) $\partial^2 W/\partial y^2$ (c) $\partial^2 W/\partial x \partial y$ (d) $\partial^2 W/\partial y \partial x$ (e) First (single) raw image of the 2D mesh scan of the speckle generator: the beam passes through the speckle generator and is reflected on the mirror before impinging onto the detector. (f) 3D rendering of the mirror's tangential slope error. The darker surface shows the slope error obtained with the ESRF Long Trace Profilometer. \label{fig:3Dmirror}}
\end{figure}

The experimental setup used corresponds to that shown in Fig.~\ref{fig:setupreflect} where the camera is kept fixed during the 2D mesh scan of the speckle generator. The method applied here was a direct and complete application of the method described in \citet{berujon2014oe} in two dimensions, which implies in practice a quadratic level of complexity. Four dimensional data from XSS is used to obtain four curvature fields that are then iteratively integrated to obtain the 2D surface profile~\cite{berujon2012ol}.

In more detail, from a 2D scan of the diffuser, four curvature fields of the wavefront $W$ are obtained by applying the formulae of XSS in a self-correlation mode. Using the notation $\mathbf{r}$ for the transverse position of a pixel on the detector (i.e. on the basis $(\mathbf{x,y})$ transverse to the beam) and $N$ being an integer, we note $\mathbf{\Delta r_x} = N p_{ix}\cdot\mathbf{x}$ or $\mathbf{\Delta r_y} = N p_{ix}\cdot\mathbf{y}$ for the distance between two pixels located on the same row or column of the detector respectively. Thus we correlate the speckle signals of pixels with the signals received in the upper and lateral neighbouring pixels located at $\mathbf{r}+\mathbf{\Delta r}$ in order to find the corresponding signals' displacement vector $\upsilon_{x,y}$. Each cross-correlation operation provides two curvature fields $\kappa_{ij}$ with $i,j \in {[x,y]}$ in the orthogonal reference $(\mathbf{x},\mathbf{y})$ coming from the relationship $$R^{-1}_{ij} = \kappa_{ij} = \frac{(1 - \frac{\upsilon_{j}}{\mathbf{\Delta r_{i}}})}{d}$$ derived from the local magnification defining $\mathbf{\Delta r}/\upsilon = R/(R-d-l)$ (see Sec. 2.2.5 of \cite{berujon2019a}), with $R$ being the wavefront curvature radius:

\begin{equation}
\begin{split}
\kappa_{xx}&=\frac{\partial^2 W}{\partial x^2}, \kappa_{xy}=\frac{\partial^2 W}{\partial y \partial x},\\
\kappa_{yy}&=\frac{\partial^2 W}{\partial y^2},\kappa_{yx}=\frac{\partial^2 W}{\partial x \partial y}
\end{split}
\end{equation}
from which the two wavefront gradients:
$$\frac{\partial W}{\partial x}~~\textrm{and}~~\frac{\partial W}{\partial y}$$
are recovered through a 2D numerical integration method~\cite{frankot1988,southwell1980,noll1978,arnison2004,berujon2015josr}. Finally a third 2D integration of $\frac{\partial W}{\partial x}$ and $\frac{\partial W}{\partial y}$ provides $W$. For an X-ray mirror with a sag of a few micrometers and when working at shallow angles, it was shown that an iterative corrective approach must be employed for an accurate surface mapping~\cite{berujon2012}. For the present 2D approach, this iterative procedure imposes necessarily the recovery of the four curvature fields, a condition not fulfilled in \citet{wang2015josr}.

Figure~\ref{fig:3Dmirror}~(a-d) shows the four curvature fields obtained using XSS in self-correlation mode for the 2D characterisation of the Zeiss substrate used in \citet{berujon2012}. The optical surface is 50~mm long, the incidence angle of the measurement was 0.13~degree, $l \approx 400$~mm and $d=1034$~mm. The speckle generator mesh scan comprised 32x32 points. In this example, the step sizes used for the scan of the membrane were different for the horizontal and vertical directions with the respective values of 0.25\,$\mu$m and 2.9\,$\mu$m (half the size of a pixel since there is no magnification in that direction). In the processing, we used $\mathbf{\Delta r} = 4 p_{ix}$, which has a value clearly larger than that of the detector point spread function, but is still small enough to preserve the resolution provided by the detector. The FReLoN camera coupled to its optics gave an effective pixel size of $p_{ix}=5.8\,\mu$m. The magnification of the X-ray optics was then used to increase the angular sensitivity $\delta \alpha  \propto \delta \tau$ through the factor $\Gamma$ (since a good choice of speckle generator step is $\delta \tau \sim p_{ix}/\Gamma$, cf. theoretical paper) and the sampling resolution on the mirror in the longitudinal direction.
Figure~\ref{fig:3Dmirror}~(e0 is the first raw image of the 2D mesh scan used and (f) is a 3D rendering of the mirror's tangential slope error from the targeted ellipse.

\subsubsection{Multilayer interference fringes\label{sec:mlmirror}}

We illustrate the use of the previous methods for this case, which is restricted to a single dimension, providing yet another investigative application. A problem often encountered when using reflective optics, either in total reflection or with reflective multilayer coatings, is the appearance in the reflected beam of intensity fringes orientated perpendicular to the photon propagation direction and orthogonal to the reflection normal axis. These fringes can be detrimental for many applications such as imaging, where they hinder the full use of the dynamic range of the detectors, and may complicate intensity normalisation processing.

\begin{figure*}
\centering{
\includegraphics[width=16.5cm]{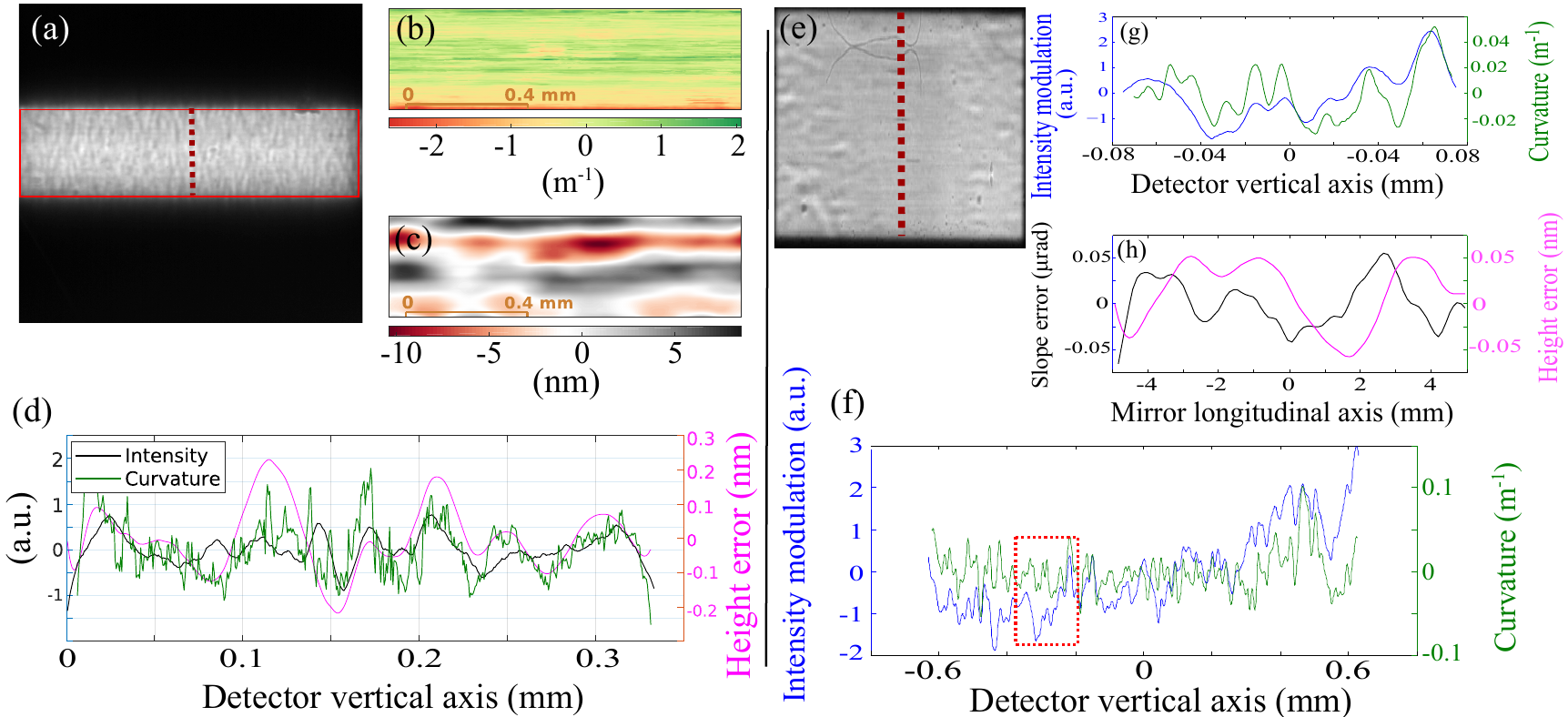}
\caption{Flat substrate metrology, left half of the figure, versus multilayer metrology, right half. \textbf{For the flat substrate:} (a) First raw image of the speckle generator upon total external reflection on the mirror. (b) Measured tangential curvature of the mirror, light being reflected in the vertical direction and falling onto the square marked in (a). Corresponding tangential (b) slope error and (c) height error. (d) Intensity (black line), curvature (purple line) and height profile (pink line) measured across the mirror along the dashed line in (a). \textbf{For the multilayer coated mirror:} (e) First raw image of the speckle generator upon reflection on the mirror surface. (f) The intensity modulation across the reflected beam marked by the dashed line in (e) is plotted in blue and the curvature measured for this same profile is plotted in green. (g) Zoom onto the region marked by the red square in (f) after the removal of a polynomial of order 2. (h) Slope and height profiles corresponding to the curvature of (g).\label{fig:MLmirror}}}
\end{figure*}

According to \citet{morawe2013}, these fringes originate mainly from the shape error of the substrate which is transferred to the conformal multilayer coating. To investigate this effect, two measurments were performed. First, an uncoated mirror was illuminated under total external reflection and later another high quality substrate after coating at the multilayer Bragg angle. The XSS method was applied in one dimension to benefit from its higher angular sensitivity and its high resolution. We present thus results of a) a 90~mm long silica substrate polished at the Institut d'Optique Graduate School and b) the sample coated with a multilayer at the ESRF with the following deposition characteristics: $\left[W/B_4C\right]_{60}$ and a $d\textrm{\small -spacing}=4$~nm.

The speckle generator was placed, as in the previous experiment, at a 400~mm distance upstream of the mirror, while the distance from the X-ray optics to the detector was $d=915$~mm and $d=955$~mm for the respective measurements. The grazing incidence angle of the photon onto the mirror was 0.21 degrees and the working energy $E=8$~keV for the uncoated substrate and $E=17$~keV for the multilayer coating. The detector used was the PCO Edge 4.2 camera coupled with optics giving a pixel size of 0.63~$\mu$m. The scan of the speckle generator consisted of 150 points of different vertically aligned positions with a step size of 0.25~$\mu$m.

Figure~\ref{fig:MLmirror} (d) shows a good correlation between the observed intensity modulation and the measured mirror curvature for the uncoated substrate. On the same graph, one can observe that the corresponding problematic height error amplitude is in the order of the wavelength, i.e. at the \r{A}ngstr\"{o}m level. In the right part of Fig.~\ref{fig:MLmirror}~(f-g), the correlation between the intensity modulation and the local wavefront curvature is even more striking while the peak-to-valley corresponding defects in the medium spatial wavelength are smaller than the \r{A}ngstr\"{o}m. This range of height error is inaccessible with current offline metrology, and currently available technology and techniques do not allow for the correction of such small defects.

Fresnel-Kirchhoff numerical propagators permit today to accurately predict the intensity produced by an optics with known errors. Yet, on-going work on the inverse problem should help us define clear specifications for the mirror design in terms of spatial spectral power. As a matter of fact, one can prove a relationship linking the intensity modulations to the mirror curvatures in the mm$^{-1}$ spatial frequency. Eventually, such mirror curvature corresponds to mirror defects' height with an amplitude in the order of the wavelength. The experimental demonstration of this relationship supports the theoretical model published by \citet{nicolas2013}, which explains the intensity modulation in term of surface curvature error. Older models with visible light \cite{Berry_2005} and whose analytical solutions can be easily applied to the X-ray regime are also consistent with the present observations.

\subsubsection{Adaptive optics optimization}

Here we apply the speckle scanning methods to X-ray adaptive optics optimization, which is, for instance, relevant in achieving smaller focal spots by actively reducing the optical aberration \cite{satoshi2016}. Although a first demonstration was described in \citet{berujonphd} where both XST and XSS were combined, further work has highlighted the advantages of the XSS technique for online X-ray optics optimization due to its high angular sensitivity.

\begin{figure}
\includegraphics[width=8.5cm]{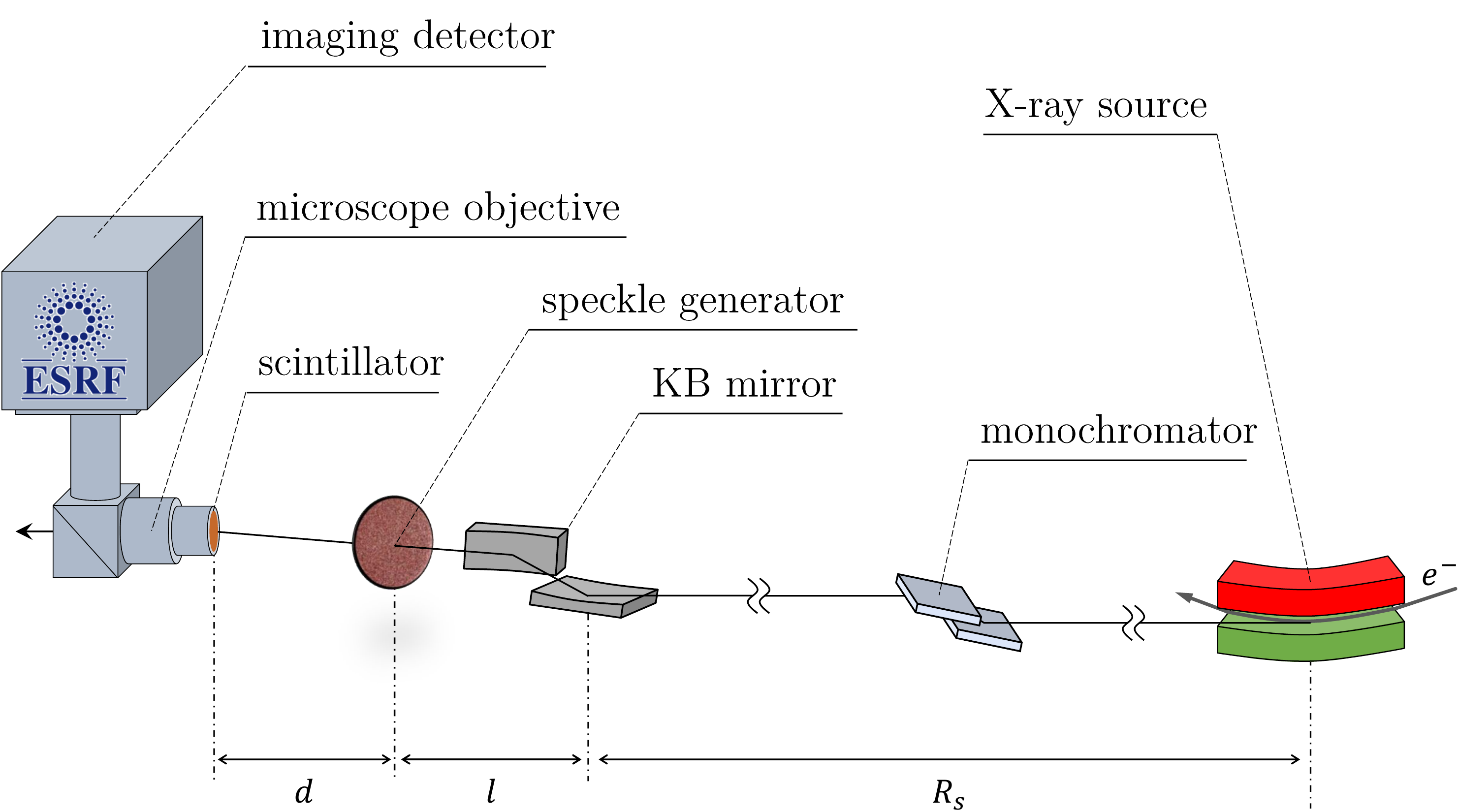}
\caption{Configuration for focusing of the KB bender.\label{fig:setupKB}}
\end{figure}

The most frequently used type of X-ray adaptive optics are deformable mirrors with controllable surfaces. These optics can be either of the bimorph kind, with a small or a large number of piezo actuators, or they can be mechanical benders that use motors with incremental steps in the nanometer or sub-nanometer range. Whilst bimorph mirrors usually offer a larger degree of freedom due to the use of more actuators, they also require electronics with high stability to keep the actuator steady over long periods of time, which is not the case with dynamically bent mirrors.

\begin{figure}
\includegraphics{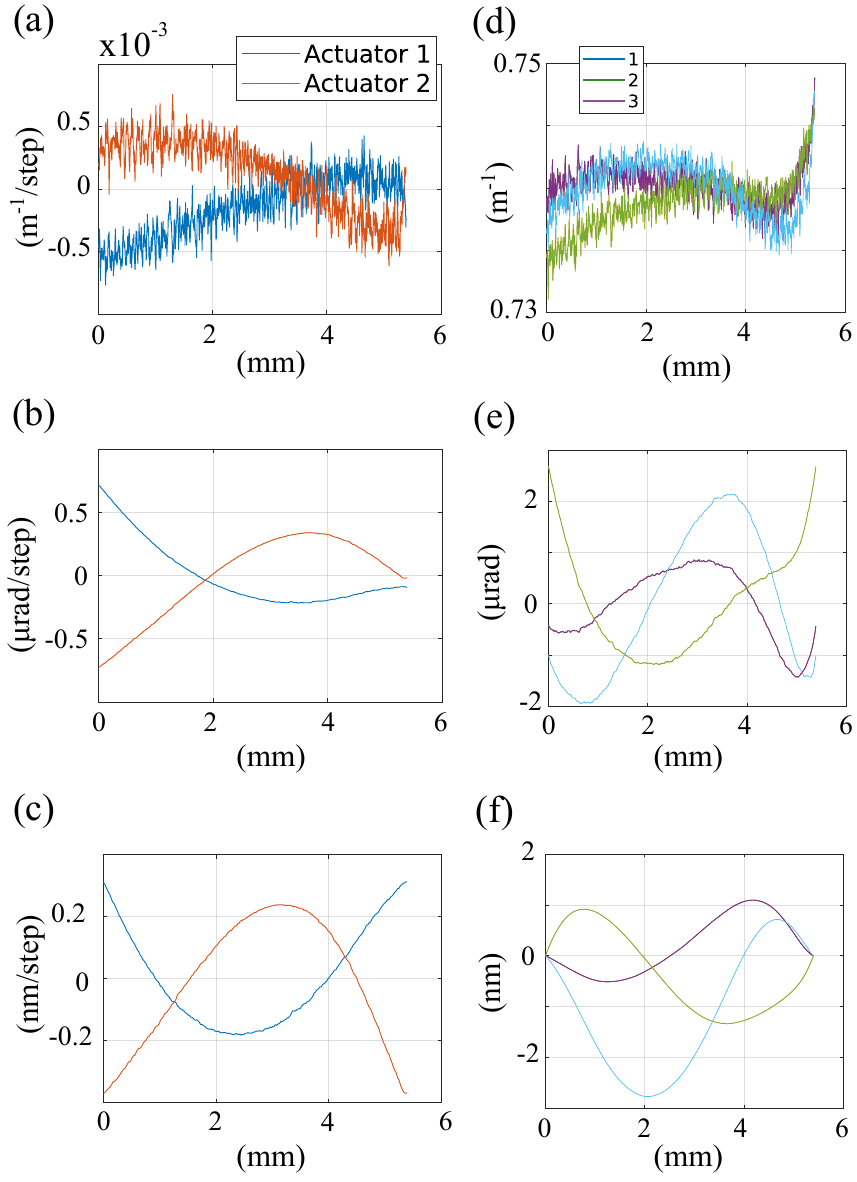}
\caption{Influence function of the actuators on the wavefront (a) curvature, (b) slope and (c) height as a function of the detector's vertical axis. Errors on the wavefront (a) curvature, (b) slope and (c) height as a function of the detector's vertical axis after the first, second and third corrections.\label{fig:ao}}
\end{figure}

The BM05 beamline of the ESRF is equipped with a Kirkpatrick-Baez reflective optics systems composed of two mechanically bendable mirrors mounted orthogonally. Each mirror bender has two encoded actuators at the ends of the mirror surface. For metrology, this system permits to evaluate various schemes for optics optimization and to explore the potential of the methods.

The setup of this experiment is shown in Fig. \ref{fig:setupKB} with the speckle generator located this time downstream of the reflective optics. Placing the speckle generator after the adaptive optics is important since we want to optimize the ensemble of all the optical components present along the beamline, hence they have to be upstream of the speckle generator. What matters here is the final state of the beam phase, adding the contribution of all the beamline optics.

A good explanation of the theory of wavefront optimization through matrix inversion and of the possible processing schemes is given in \cite{tyson2010}. In short, wavefront optimization through matrix inversion consists of a first phase of measuring, for each of the actuators, the effect of an incremental step on the wavefront or its derivatives. Hence, differential metrology is required for this task. It can be done either by working in a differential mode as described earlier, or by using the accurate and easy to implement XSS self-correlation mode, as is the case here. This consists of subtracting the wavefront measured before and after having applied a command to an actuator. The response function of each actuator is then placed as a column of a matrix $A$ with as many columns as the number of degrees of freedom (actuators) that the system contains. This process implies that the system is repeatable and even linear, which often calls for encoders to guarantee a correct displacement of each actuator.

Then, by using the notation $\Upsilon$ for the vector wavefront error (or respectively for its gradient or curvature) as measured with absolute metrology, and $\Psi$ for the piezo correction that needs to be applied, the problem requires solution of the linear equation $\Upsilon + A\Psi=0$. This equation is usually solved in the least square sense by matrix inversion to obtain $\Psi = -A^{-1}\Upsilon$. Because the matrix A is not square and singular with computer working precision, the inversion is operated in the Moore-Penrose pseudo-inverse sense using singular value decomposition, Gauss elimination, Cholesky decomposition or any other alternative numerical method \cite{barnett90}.

The optimization of the wavefront can be done on the wavefront surface, on its gradient or on the curvature, all three being linked through an integral relation. Usually, this question can be answered by the nature of deformation created by the actuators: bimorph mirrors combine well with curvature sensing systems while benders provide better results when used in combination with slope measurements \cite{tyson2010}. Regarding the fitting of polynomials to the wavefront error, one can also fit the derivative of the surface polynomial to the slope error to directly infer the wavefront error in a modal manner. Even so, in the X-ray regime, the zonal approach can be favoured since the resolutions used are high with respect to what is encountered in visible light optics. Such a zonal approach permits a full description of the system, although only the low orders can be compensated.

Figure~\ref{fig:ao} shows the actuator influence function on the (c) wavefront, (b) wavefront slope and (a) wavefront curvature of the two actuators of the BM05 vertical focusing bender, measured at 240~mm from the mirror focus (which for this case was also set to approximately one focal distance of the mirror for convenience). Note that the tilt has been removed by subtracting the mean value of the slope. Figures~\ref{fig:ao} (d-f) show the wavefront errors, the wavefront gradient errors and curvature errors after three correcting iterations. One can observe that the method converges quickly thanks to the high accuracy of the online metrology, while the limited number of actuators of the system is the main limitation on the system's performance. Indeed, the S-shape of the remaining wavefront error, corresponding to a polynomial of order three, indicates that more than two actuators would be needed to further correct the optics. Further work on bimorph mirrors with more actuators proved the validity of the concept for higher order systems \cite{berujonphd}.

\subsection{Refractive optics}

\subsubsection{Single compound refractive lens}

Another interesting application of X-ray speckle-based metrology is the characterisation of inline refractive optics, especially to assess the shape errors of lenses prior to permanent installation on a beamline.

\begin{figure}
\includegraphics[width=8.5cm]{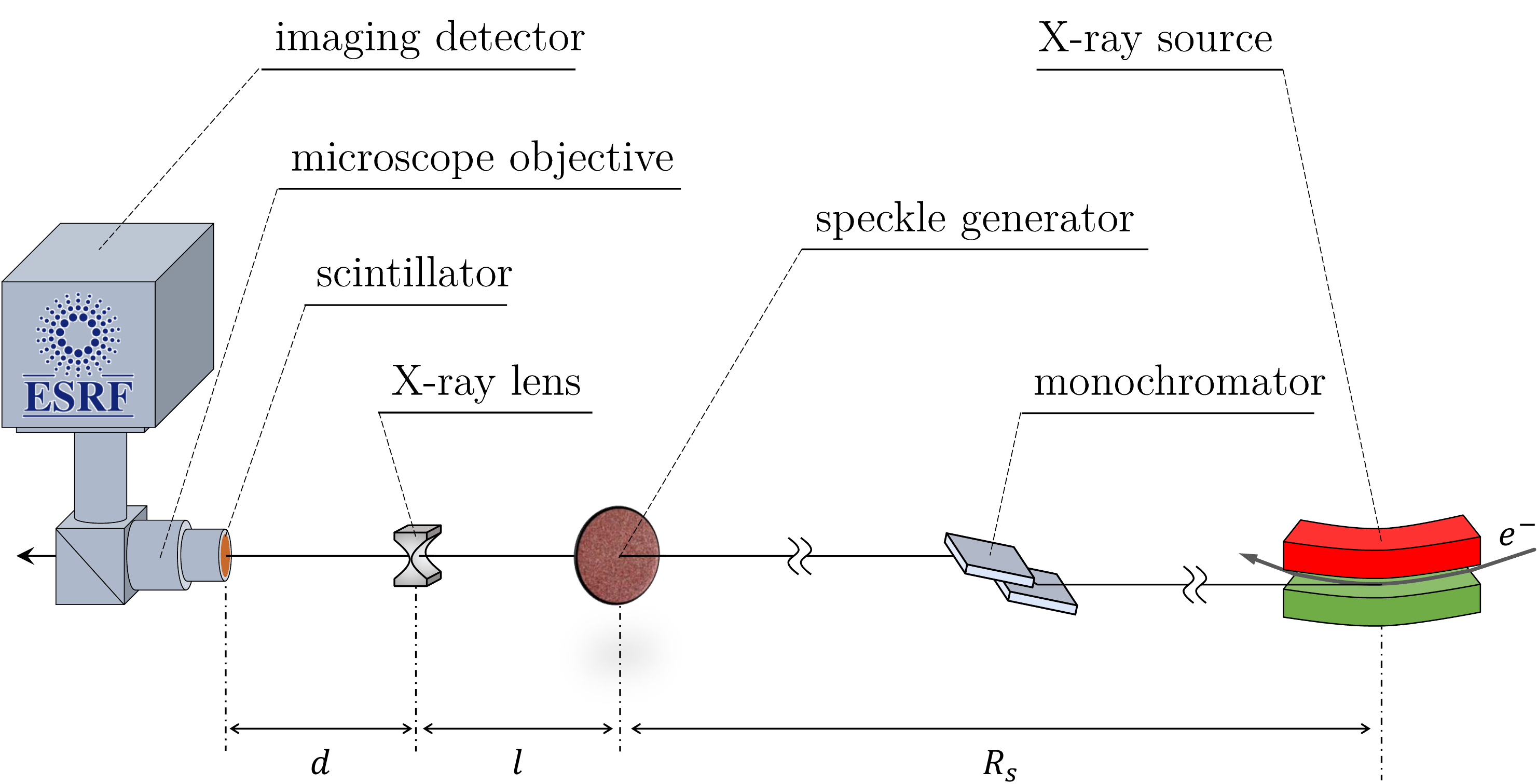}
\caption{Setup for refractive optics characterisations.\label{fig:setuprefract}}
\end{figure}

For such at-wavelength characterisation, we use the setup sketched in Fig.~\ref{fig:setuprefract}, where the speckle generator, the optical component, and the detector are all aligned. XST, XSVT, XSS or any more advanced scheme can be employed here to recover the phase shift induced by an overall weakly (single lens) refractive optics. We choose to implement the XSVT technique for routine lens characterisation at the ESRF, due to the following points:
\begin{itemize}
  \item We want to keep the setup simple for such characterisations and a probe beam left with its original collimation (no additional focusing or beam shaping optics are used). The lateral resolution of XSVT is the detector resolution, which is higher than the one obtained using the XST technique (averaging over pixel groups). Besides, the XSVT resolution equals the one achieved with the other remaining schemes which however do not provide additional sensitivity benefits when using an almost collimated probe beam.
  \item The angular dynamical range of the XSVT technique is theoretically not limited as in the case of XSS, where the size of the speckle generator mesh scans sets the maximum displacement vector and hence the phase gradient. This aspect is of value when measuring a series of lenses with varying design parameters.
  \item The number of necessary images for using properly the XSVT is in the order of $2\times N$ images taken at random positions of the speckle generator while it is in the order of $M^2$ images for XSS and hybrid schemes (M being the number of lateral points in the membrane position mesh and typically in the range $[11;31]$). In the following case we use $N\!=\!81$ images in each scan with the speckle generator located at 81 positions on a $\log_2$ grid. The reference data set can be reused for several lens measurements. At the ESRF, lenses are often measured in batches of 10 which corresponds to the collection of 891 images, or 972 when using two different reference data sets to evaluate and compensate for eventual beam drifts and for fluctuations of the beamline.
\end{itemize}

Here and in Sec.~\ref{sec:monoreq}, we apply the technique in differential mode. This means that two stacks of images are used to analyse each lens: one with the lens present in the beam and a reference data set acquired without the lens. Using the differential mode permits to isolate the contribution of the investigated optics (the lens) on the X-ray beam phase, putting aside the errors coming from the upstream optics.

\begin{figure}
\includegraphics[width=8.5cm]{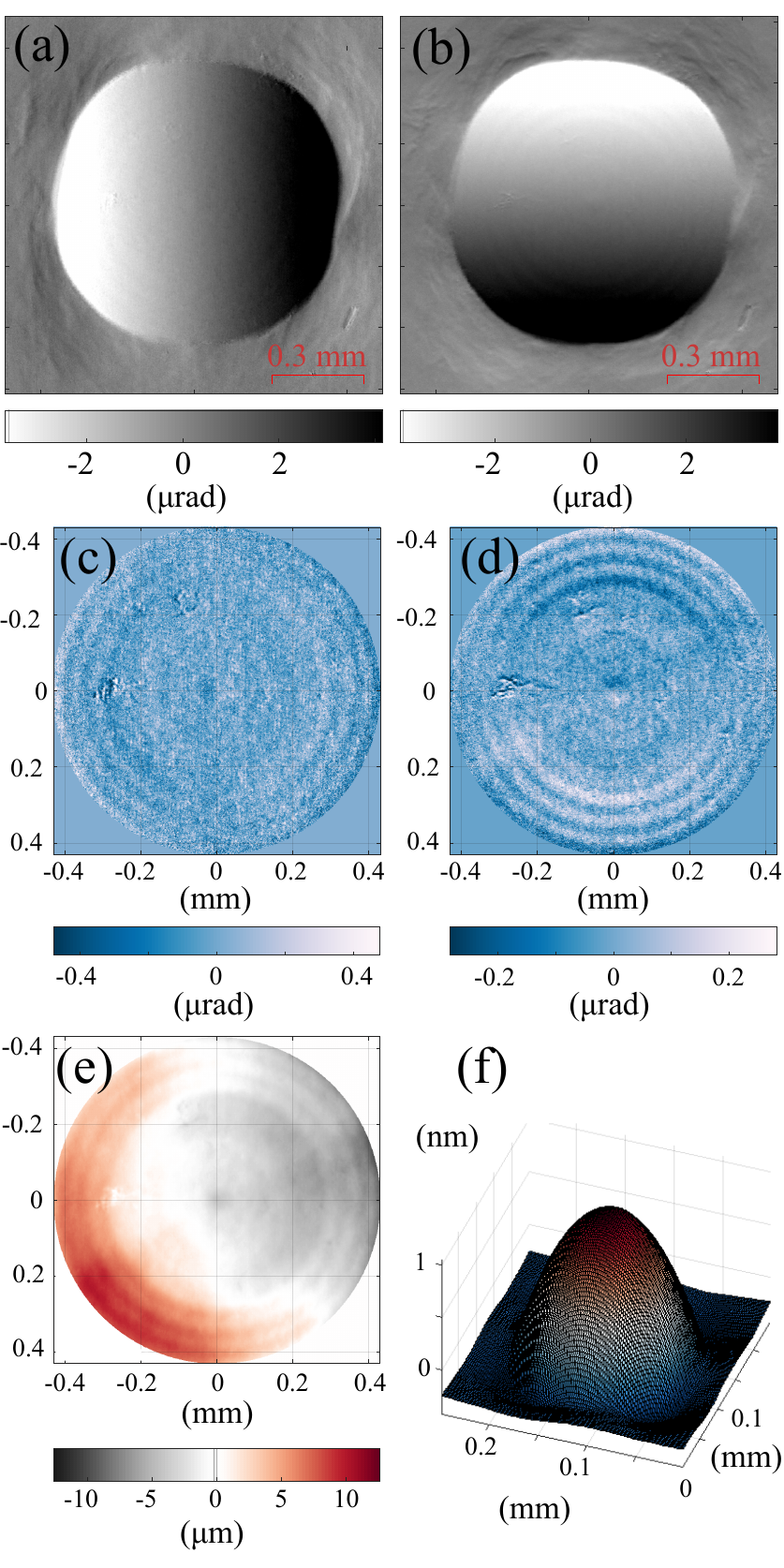}
\caption{(a) Horizontal and (b) vertical wavefront gradients. (c) Horizontal and (d) vertical wavefront gradient errors. (e) Projected shape error recovered from the (f) measured wavefront.\label{fig:lens}}
\end{figure}

Figure~\ref{fig:lens} shows the information retrieved for an aluminium lens using the XSVT techniques from two stacks of images. The lens was produced in a first batch of Al lenses at the ESRF with a design radius at the apex of $R_{apex}=50~\mu$m and a physical aperture of $\sim 440~\mu$m. The speckle generator was made of cellulose acetate with a 1.2~$\mu$m pore size. The distance from the speckle generator to the lens was 400~mm and the distance from the lens to the detector was $d=905$~mm. The detector was a PCO Edge 4.2 camera coupled to a 10x microscope objective. From the undistortion step of the images, the effective pixel of the system was calculated as $p_{ix}=0.63~\mu$m. All 2D maps shown in Fig.~\ref{fig:lens} have been corrected for detector distortion. The photon energy was $E=17$~keV.


The differential wavefront gradients induced by the lens (Fig.~\ref{fig:lens} (a-b)) were computed using the XSVT correlation equation presented in the theoretical paper. The wavefront gradient errors were calculated by removing the best vertical, respectively horizontal, plane fit from the wavefront gradients. The wavefront error $W_\textrm{err}$ of the lens was obtained by integration of the wavefront gradient errors and finally the thickness error $T_\textrm{err}$ was found by the linear relation:
$T_\textrm{err}=W_\textrm{err}/\delta(\lambda)$. For the case presented here $\delta_{Al}(17~\mathrm{keV}) = 1.87\times 10^{-6}$.

Assuming the wavefront being the sum of a perfect spherical surface $W_s$ and an error deviation from it $W_\textrm{err}$, we can calculate a differential focusing distance $f$ along an axis $\hat{\mathbf{r}}=\mathbf{r}/|\mathbf{r}|$ from a two dimensional fit of the surface derivative of $W = W_s+W_\textrm{err}$:
\begin{equation}
f =\Bigg \langle\frac{\partial^2 W}{\partial \hat{\mathbf{r}}^2} \Bigg\rangle
 = \Bigg\langle\frac{\partial^2 W_s}{\partial \hat{\mathbf{r}}^2}\Bigg\rangle
\end{equation}
where $\langle.\rangle$ denotes the quadratic mean and by definition we have $\langle W_{err}\rangle=0$. Note that for a perfectly rotationally symmetrical lens, $f$ is constant for any $\mathbf{r}$. Experimentally this is however often not the case, with most lenses presenting some degree of astigmatism.

The calculation of $f$ permits determining the radius of a single bi-concave lens at the apex, which is equal to
\begin{equation}
R = 2f\delta(\lambda)
\end{equation}
For the lenses presented, one can observe in Fig.~\ref{fig:lens} (c-e) rings which are attributed to imperfections in the machining of the punches that have been used during the lens fabrication process.

This XSVT method is now routinely applied at the ESRF for the inspection of newly acquired or manufactured lenses. With only a few minutes necessary for the characterisation of a lens, the very sensitive method is easy to implement for batch analysis. Stacks of lenses are also studied following the same methodology. In that case, usually a higher energy is selected. This allows the detector to intercept a beam that is still large enough - at lower energies the focusing would reduce the beam size at the detector - and thus provides an acceptable resolution.

This type of metrology data represents what is called a zonal approach, i.e. 2D pixelized maps of the optics aberrations. From these, a modal decomposition can be obtained \cite{tysonbook}. Indeed, an easy and fast way of representing a lens shape error consists of decomposing its resulting aberrated wavefront onto orthonormal polynomials \cite{mahajan2007}. This description permits a more rapid interpretation of the type of defects of the lens and allows to compare it to other optics as is routinely done for visible light optics.

\subsubsection{Strongly refracting optics\label{sec:strefop}}

The previously studied Al single lens is weakly focusing, as is the case for most refractive optics in the X-ray regime \cite{roth2017}. Yet, some X-ray refractive optics consist of several elements being printed or made from a single bulk, which cannot be dismantled into elementary parts, thus presenting a much stronger focusing effect.

\begin{figure}
\includegraphics[width=8.5cm]{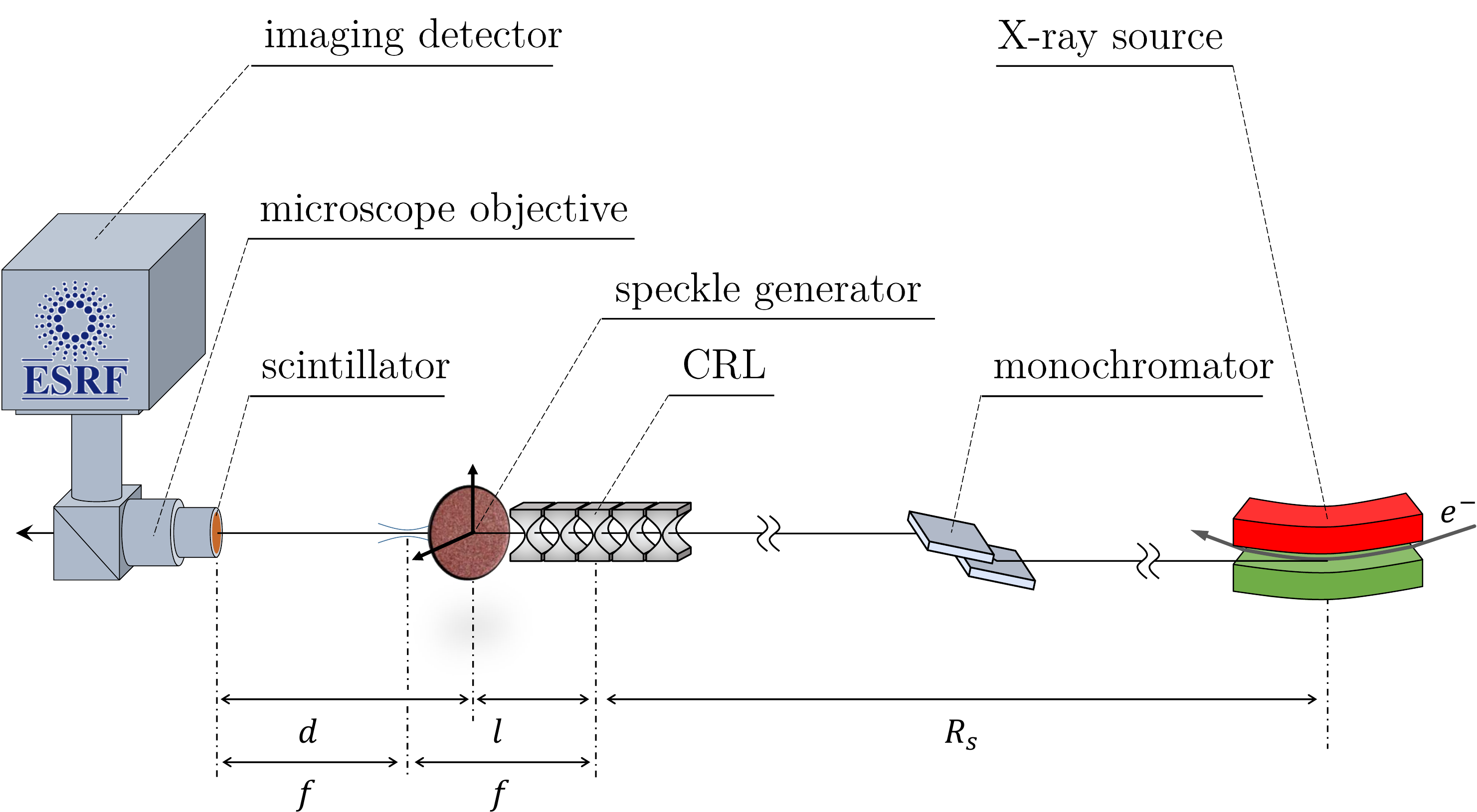}
\includegraphics[width=8.5cm]{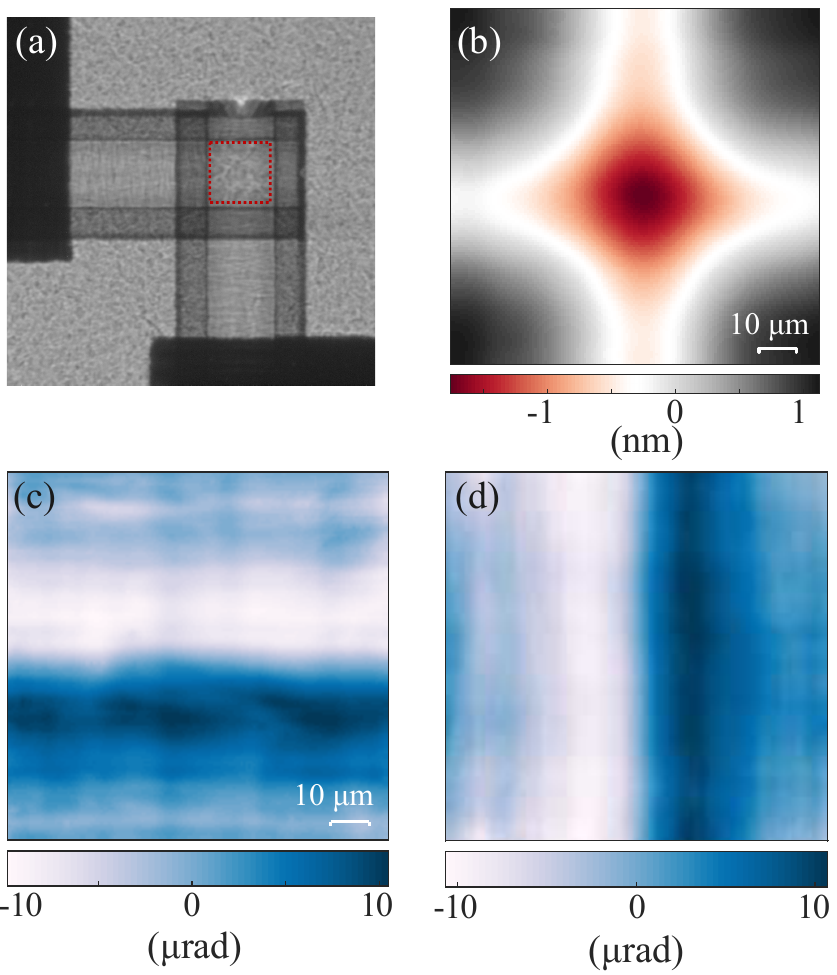}
\caption{Top: Setup for the analysis of a strongly focusing lens stack. (a) First image of the stack showing the crossed lenses with the overlaying randomly modulated wavefront. (b) Calculated wavefront error for the area marked with the inset in (a). (c) Vertical and (d) horizontal wavefront gradient errors of the same area.\label{fig:lensabs}}
\end{figure}

When the focusing effect is very strong and/or when the phase distortion induced varies rapidly across the aperture, the implementation of a speckle technique in the differential mode, as done in the previous section, can become difficult or impossible. Indeed, with a very large and spatially rapidly varying distortion of the wavefront, the correlation algorithm can no longer track the modulation pattern from one stack of images to the other. In that case, the absolute mode can become efficient in the hypothesis that the phase defects present in the incoming beam are much smaller than those generated by the optics under investigation. For a beamline such as BM05, where only two Beryllium windows and two flat crystals interact with the beam (with no other focusing optics), this assumption is often valid.

The XSS self correlation mode (see Sec. 2.2.5 of the theoretical paper) can be employed in the same way as for the online characterisation of the mirror in Sec.~\ref{sec:2dmirror} and \ref{sec:mlmirror}. Figure~\ref{fig:lensabs} illustrates the method on a SU-8 lens stack consisting of two crossed 1D lens stacks produced by X-ray LIGA.

\begin{figure*}
\centering{
\includegraphics[width=16.5cm]{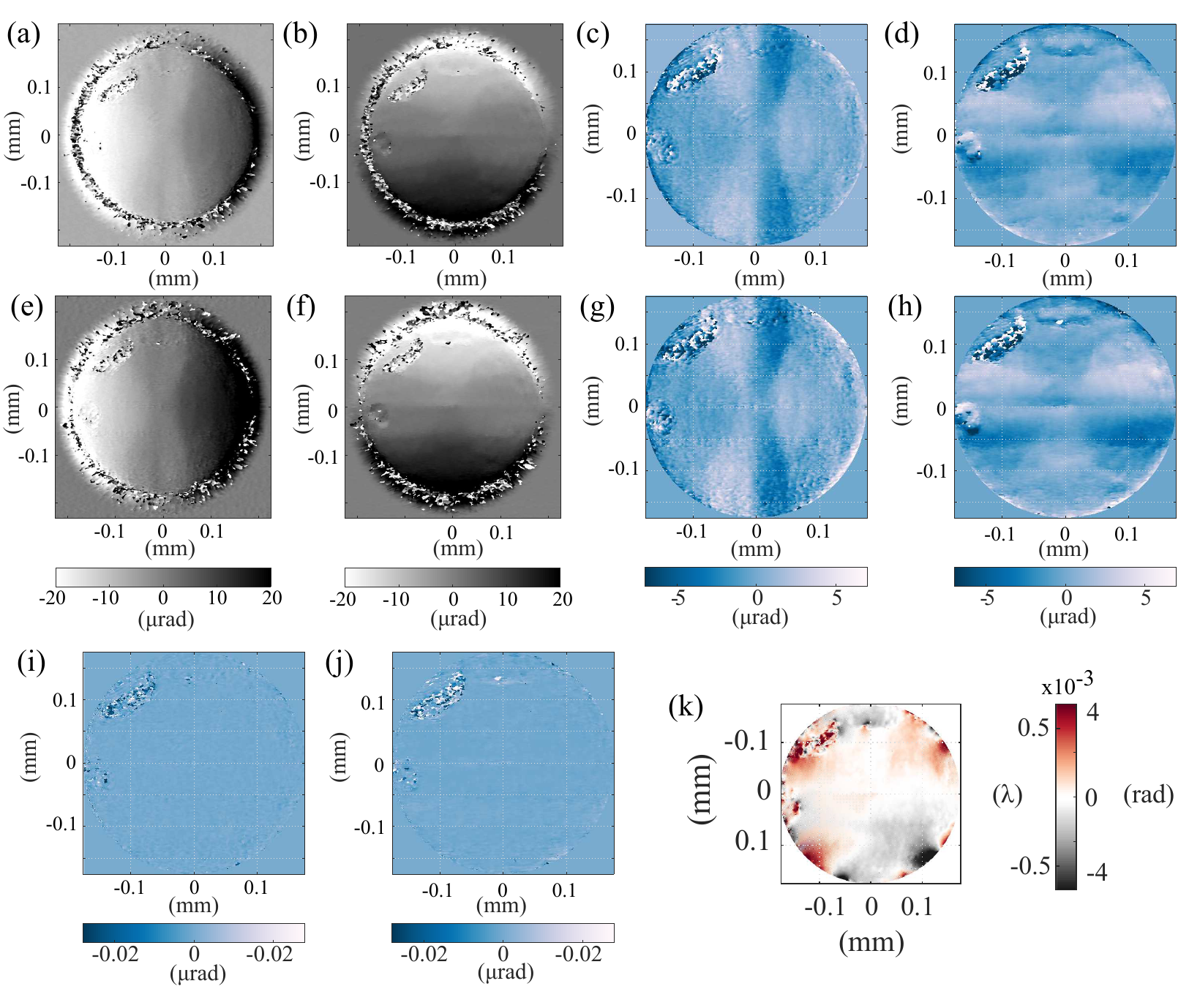}
\caption{Diamond lens prototype characterisation. (a) Horizontal and (b) vertical differential wavefront gradients measured using a double crystal monochromator and (c-d) associated wavefront gradient error. (e-h) Equivalent measurements obtained using a multilayer monochromator. (i) Difference map of the horizontal gradients and (j) of the vertical gradients. (k) Difference phase map calculated using (i-j).\label{fig:compDCMDMM}}}
\end{figure*}

The focal length of this lens, comprising 65 vertically aligned lenslets and 65 horizontally aligned lenslets, at the measurement energy of $E$~=~30~keV, was 250~mm. Note that the speckle generator was mounted after the lens, at a distance $l = 95$~mm. This location of the speckle generator overcomes the problems encountered when the lens material scatters too strongly thus blurring or masking the random modulation pattern. The detector was located at $d = 405$~mm from the speckle generator or $d = 2f$. At this position, the magnification of the random speckle pattern is equal to about -1. The speckle generator was scanned twice ($2\times 100$ images), once in each transverse direction of the beam with a step size of 1~$\mu$m. As in Sec.~\ref{sec:2dmirror}, the wavefront gradient errors were recovered from integration of the curvature maps obtained by the XSS technique where $\kappa = (1 - \upsilon/\mathbf{\Delta r})/d$ (see \cite{berujon2019a}). Displacement vectors were calculated from pixels located at a distance of $\mathbf{\Delta r} = 4 p_{ix}$. Figure.~\ref{fig:lensabs} shows that this lens stack suffers from significant wavefront gradient errors of the order of a few microradians. This is about 30 to 40 times higher as seen for a single Al lens in Fig.~\ref{fig:lens}. However, here we have a stack of $2\times 65$ lenses, which means that the error on an individual lens basis is of the same order of magnitude, if not smaller.


\subsubsection{Monochromaticity requirements \label{sec:monoreq}}

The lens analysis examples shown above were obtained under monochromatic conditions with an energy selectivity corresponding to $\Delta E /E\sim 10^{-4}$. For a quick comparative study, a diamond lens known for having noticeable aberrations was measured in similar conditions but with two different bandwidths, of $\Delta E /E\sim 10^{-4}$ as before and of $\Delta E /E\sim 10^{-2}$ using a multilayer monochromator. In both cases, the mean photon energy was $E=15~\mathrm{keV}$.

The lens was characterised in both cases using the XSVT technique with 49 scanning points (cf. Sec.~2.2.3 in \cite{berujon2019a}) to recover the wavefront gradient and shape. The comparative results presented in Fig.~\ref{fig:compDCMDMM} show that these results are in very good agreement despite a spectral bandwidth ratio of $\sim 70$. The optical aberrations of the lens shown in (c-d) and (g-h) match very well down to the sub-microradian scale. This can be more precisely observed in the wavefront gradient difference maps shown in (i-j). Therein, the gradient differences are homogeneous and in the range of tens of nanoradians except at points where the correlation was poor due to a lack of light. Finally, the difference phase map shown in (k), calculated from the maps (i)-(j), confirms the good agreement between the two measurements, with a wavefront difference between the two measurements smaller than 0.001 wavelength.

This experiment quantitatively confirms that in the near-field, requirements on the longitudinal monochromaticity are not stringent. Whilst a bandwidth of a couple of percent is still very selective with respect to a laboratory source, numerical simulations showed that by working with a broader source spectrum and introducing the concept of equivalent energy, quantitative results were still achievable \cite{zdora2015}. However, for quantitative metrology, the refractive index of the material $\delta(\lambda_{eq})$ must be carefully selected in order to recover the material's thickness.

\section{Conclusions} 


Several examples employing X-ray speckle-based methods for the characterisation of X-ray optics were presented, illustrating the potential of this approach in a variety of situations.
The applicability of X-ray speckle methods has been demonstrated with a low coherence source. This suggests that the presented schemes could possibly be applicable to laboratory sources, provided a few adjustments are implemented, especially in term of beam monochromaticity. These at-wavelength characterisation methods are now routinely available at synchrotrons as illustrated in this paper. They may also become attractive for the XFEL sources that have recently started operation. Speckle based methods should thus help in the manufacturing of better optics during their production stage or, for instance, enable the manufacturing of compensating optics.

Finally, while the present paper reviews some fundamental concepts in using speckle based methods for adaptive optics, we believe that particularly the XSS method allows to achieve the ultimate optimization and focusing of x-rays with actuator optics and will be helpful on beamline to achieve easily and routinely the optimal alignments of nanobeam optics.

\vspace{0.5cm}

The authors wish to thank the ESRF for financial and personal support. The authors are grateful to S. Antipov and S. Stoupin for the loan of the diamond lens, T. Manning for the production of Al lenses, A. Vivo for the offline metrology, the Institut d'Optique Graduate School for the production and loan of a polished substrate, C. Morawe for multilayer deposition and, C. Detlefs and A. Last for the loan of the SU-8 crossed lens stacks.


%

\end{document}